%% file: NuEXSPaper.tex
\RequirePackage{lineno}
\documentclass[%
 reprint,
 superscriptaddress,
showpacs,preprintnumbers,
amsmath,
amssymb,
aps,
prl,
floatfix,
]{revtex4-1}

\usepackage{graphicx}
\usepackage{epstopdf}
\usepackage{dcolumn}
\usepackage{bm}
\usepackage{color}
\usepackage{xspace}
\usepackage{hyperref}
\usepackage{natbib}
\usepackage{multirow}
\input symbols
\linenumbersep{0.1cm}

\begin{document}



\title{Measurement of the Inclusive Electron Neutrino Charged Current Cross Section on Carbon with the T2K Near Detector}

\input{author-jun03_2014}


\date{\today}

\begin{abstract}
\input{abstract}
\end{abstract}
\maketitle

\input{intro}

\input{T2K-setup}

\input{neutrino_beam}

\input{data-selection}

\input{unfolding}

\input{results}

\input{conclusion}

\input{acknowledge}

\bibliographystyle{apsrev4-1}
\bibliography{NuEXSPaper}

\end{document}

%% file: author-jun03_2014.tex
 
\newcommand{\INSTC}{\affiliation{University of Alberta, Centre for Particle Physics, Department of Physics, Edmonton, Alberta, Canada}}
\newcommand{\INSTEE}{\affiliation{University of Bern, Albert Einstein Center for Fundamental Physics, Laboratory for High Energy Physics (LHEP), Bern, Switzerland}}
\newcommand{\INSTFE}{\affiliation{Boston University, Department of Physics, Boston, Massachusetts, U.S.A.}}
\newcommand{\INSTD}{\affiliation{University of British Columbia, Department of Physics and Astronomy, Vancouver, British Columbia, Canada}}
\newcommand{\INSTGA}{\affiliation{University of California, Irvine, Department of Physics and Astronomy, Irvine, California, U.S.A.}}
\newcommand{\INSTI}{\affiliation{IRFU, CEA Saclay, Gif-sur-Yvette, France}}
\newcommand{\INSTGB}{\affiliation{University of Colorado at Boulder, Department of Physics, Boulder, Colorado, U.S.A.}}
\newcommand{\INSTFG}{\affiliation{Colorado State University, Department of Physics, Fort Collins, Colorado, U.S.A.}}
\newcommand{\INSTFH}{\affiliation{Duke University, Department of Physics, Durham, North Carolina, U.S.A.}}
\newcommand{\INSTBA}{\affiliation{Ecole Polytechnique, IN2P3-CNRS, Laboratoire Leprince-Ringuet, Palaiseau, France }}
\newcommand{\INSTEF}{\affiliation{ETH Zurich, Institute for Particle Physics, Zurich, Switzerland}}
\newcommand{\INSTEG}{\affiliation{University of Geneva, Section de Physique, DPNC, Geneva, Switzerland}}
\newcommand{\INSTDG}{\affiliation{H. Niewodniczanski Institute of Nuclear Physics PAN, Cracow, Poland}}
\newcommand{\INSTCB}{\affiliation{High Energy Accelerator Research Organization (KEK), Tsukuba, Ibaraki, Japan}}
\newcommand{\INSTED}{\affiliation{Institut de Fisica d'Altes Energies (IFAE), Bellaterra (Barcelona), Spain}}
\newcommand{\INSTEC}{\affiliation{IFIC (CSIC \& University of Valencia), Valencia, Spain}}
\newcommand{\INSTEI}{\affiliation{Imperial College London, Department of Physics, London, United Kingdom}}
\newcommand{\INSTGF}{\affiliation{INFN Sezione di Bari and Universit\`a e Politecnico di Bari, Dipartimento Interuniversitario di Fisica, Bari, Italy}}
\newcommand{\INSTBE}{\affiliation{INFN Sezione di Napoli and Universit\`a di Napoli, Dipartimento di Fisica, Napoli, Italy}}
\newcommand{\INSTBF}{\affiliation{INFN Sezione di Padova and Universit\`a di Padova, Dipartimento di Fisica, Padova, Italy}}
\newcommand{\INSTBD}{\affiliation{INFN Sezione di Roma and Universit\`a di Roma ``La Sapienza'', Roma, Italy}}
\newcommand{\INSTEB}{\affiliation{Institute for Nuclear Research of the Russian Academy of Sciences, Moscow, Russia}}
\newcommand{\INSTHA}{\affiliation{Kavli Institute for the Physics and Mathematics of the Universe (WPI), Todai Institutes for Advanced Study, University of Tokyo, Kashiwa, Chiba, Japan}}
\newcommand{\INSTCC}{\affiliation{Kobe University, Kobe, Japan}}
\newcommand{\INSTCD}{\affiliation{Kyoto University, Department of Physics, Kyoto, Japan}}
\newcommand{\INSTEJ}{\affiliation{Lancaster University, Physics Department, Lancaster, United Kingdom}}
\newcommand{\INSTFC}{\affiliation{University of Liverpool, Department of Physics, Liverpool, United Kingdom}}
\newcommand{\INSTFI}{\affiliation{Louisiana State University, Department of Physics and Astronomy, Baton Rouge, Louisiana, U.S.A.}}
\newcommand{\INSTJ}{\affiliation{Universit\'e de Lyon, Universit\'e Claude Bernard Lyon 1, IPN Lyon (IN2P3), Villeurbanne, France}}
\newcommand{\INSTCE}{\affiliation{Miyagi University of Education, Department of Physics, Sendai, Japan}}
\newcommand{\INSTDF}{\affiliation{National Centre for Nuclear Research, Warsaw, Poland}}
\newcommand{\INSTFJ}{\affiliation{State University of New York at Stony Brook, Department of Physics and Astronomy, Stony Brook, New York, U.S.A.}}
\newcommand{\INSTGJ}{\affiliation{Okayama University, Department of Physics, Okayama, Japan}}
\newcommand{\INSTCF}{\affiliation{Osaka City University, Department of Physics, Osaka, Japan}}
\newcommand{\INSTGG}{\affiliation{Oxford University, Department of Physics, Oxford, United Kingdom}}
\newcommand{\INSTBB}{\affiliation{UPMC, Universit\'e Paris Diderot, CNRS/IN2P3, Laboratoire de Physique Nucl\'eaire et de Hautes Energies (LPNHE), Paris, France}}
\newcommand{\INSTGC}{\affiliation{University of Pittsburgh, Department of Physics and Astronomy, Pittsburgh, Pennsylvania, U.S.A.}}
\newcommand{\INSTFA}{\affiliation{Queen Mary University of London, School of Physics and Astronomy, London, United Kingdom}}
\newcommand{\INSTE}{\affiliation{University of Regina, Department of Physics, Regina, Saskatchewan, Canada}}
\newcommand{\INSTGD}{\affiliation{University of Rochester, Department of Physics and Astronomy, Rochester, New York, U.S.A.}}
\newcommand{\INSTBC}{\affiliation{RWTH Aachen University, III. Physikalisches Institut, Aachen, Germany}}
\newcommand{\INSTFB}{\affiliation{University of Sheffield, Department of Physics and Astronomy, Sheffield, United Kingdom}}
\newcommand{\INSTDI}{\affiliation{University of Silesia, Institute of Physics, Katowice, Poland}}
\newcommand{\INSTEH}{\affiliation{STFC, Rutherford Appleton Laboratory, Harwell Oxford,  and  Daresbury Laboratory, Warrington, United Kingdom}}
\newcommand{\INSTCH}{\affiliation{University of Tokyo, Department of Physics, Tokyo, Japan}}
\newcommand{\INSTBJ}{\affiliation{University of Tokyo, Institute for Cosmic Ray Research, Kamioka Observatory, Kamioka, Japan}}
\newcommand{\INSTCG}{\affiliation{University of Tokyo, Institute for Cosmic Ray Research, Research Center for Cosmic Neutrinos, Kashiwa, Japan}}
\newcommand{\INSTGI}{\affiliation{Tokyo Metropolitan University, Department of Physics, Tokyo, Japan}}
\newcommand{\INSTF}{\affiliation{University of Toronto, Department of Physics, Toronto, Ontario, Canada}}
\newcommand{\INSTB}{\affiliation{TRIUMF, Vancouver, British Columbia, Canada}}
\newcommand{\INSTG}{\affiliation{University of Victoria, Department of Physics and Astronomy, Victoria, British Columbia, Canada}}
\newcommand{\INSTDJ}{\affiliation{University of Warsaw, Faculty of Physics, Warsaw, Poland}}
\newcommand{\INSTDH}{\affiliation{Warsaw University of Technology, Institute of Radioelectronics, Warsaw, Poland}}
\newcommand{\INSTFD}{\affiliation{University of Warwick, Department of Physics, Coventry, United Kingdom}}
\newcommand{\INSTGE}{\affiliation{University of Washington, Department of Physics, Seattle, Washington, U.S.A.}}
\newcommand{\INSTGH}{\affiliation{University of Winnipeg, Department of Physics, Winnipeg, Manitoba, Canada}}
\newcommand{\INSTEA}{\affiliation{Wroclaw University, Faculty of Physics and Astronomy, Wroclaw, Poland}}
\newcommand{\INSTH}{\affiliation{York University, Department of Physics and Astronomy, Toronto, Ontario, Canada}}
 
\INSTC
\INSTEE
\INSTFE
\INSTD
\INSTGA
\INSTI
\INSTGB
\INSTFG
\INSTFH
\INSTBA
\INSTEF
\INSTEG
\INSTDG
\INSTCB
\INSTED
\INSTEC
\INSTEI
\INSTGF
\INSTBE
\INSTBF
\INSTBD
\INSTEB
\INSTHA
\INSTCC
\INSTCD
\INSTEJ
\INSTFC
\INSTFI
\INSTJ
\INSTCE
\INSTDF
\INSTFJ
\INSTGJ
\INSTCF
\INSTGG
\INSTBB
\INSTGC
\INSTFA
\INSTE
\INSTGD
\INSTBC
\INSTFB
\INSTDI
\INSTEH
\INSTCH
\INSTBJ
\INSTCG
\INSTGI
\INSTF
\INSTB
\INSTG
\INSTDJ
\INSTDH
\INSTFD
\INSTGE
\INSTGH
\INSTEA
\INSTH
 
\author{K.\,Abe}\INSTBJ
\author{J.\,Adam}\INSTFJ
\author{H.\,Aihara}\INSTCH\INSTHA
\author{T.\,Akiri}\INSTFH
\author{C.\,Andreopoulos}\INSTEH
\author{S.\,Aoki}\INSTCC
\author{A.\,Ariga}\INSTEE
\author{S.\,Assylbekov}\INSTFG
\author{D.\,Autiero}\INSTJ
\author{M.\,Barbi}\INSTE
% ...
 
\collaboration{The T2K Collaboration}\noaffiliation

%% file: abstract.tex
The T2K off-axis near detector, ND280, is used to make the first differential
cross-section measurements of electron neutrino charged current interactions at
energies ${\sim}1$~GeV as a function of electron momentum, electron scattering angle and four-momentum transfer of the interaction. The total 
flux-averaged \nue charged current cross-section on carbon is measured to be 
$\fluxav=1.11\pm0.09~(\mathrm{stat})\pm0.18~(\mathrm{syst})\times10^{-38}~\mathrm{cm}^{2}/\mathrm{nucleon}$. 
The differential and total cross-section measurements agree with the predictions of 
two leading neutrino interaction generators, NEUT and GENIE. The NEUT prediction is 
$1.23\times10^{-38}~\mathrm{cm}^{2}/\mathrm{nucleon}$ and the GENIE prediction is 
$1.08\times10^{-38}~\mathrm{cm}^{2}/\mathrm{nucleon}$. The total \nue charged current cross-section result is also in 
agreement with data from the Gargamelle experiment.

%% file: intro.tex
{\it Introduction}\textemdash
T2K is a long baseline neutrino oscillation experiment measuring \nue appearance and
\num disappearance from a \num beam. Neutrino oscillations are described by a mixing
matrix parametrized by three mixing angles and a CP
violating phase, \dcp~\cite{Pontecorvo57,MNS62}.
The three mixing angles have been measured to better than 10\%
precision~\cite{PhysRevD.86.010001}, and measuring \dcp is currently a major goal in
neutrino physics~\cite{Abe:2013hdq}.

Future \nue appearance measurements can be used to search for CP violation in
neutrino interactions, and these rely on precise understanding of both \num and \nue
charged-current (CC) interaction
cross-sections at energies ${\sim}1$~GeV. Many \num cross-section measurements have
been made at the GeV scale, both of the total CC inclusive cross-section and of
individual interaction modes (see Ref.~\cite{RevModPhys.84.1307} for a review of
cross-section data, and Refs.~\cite{CCinc, Tice:2014pgu, Fiorentini:2013ezn} for
recent results). Only the Gargamelle experiment has measured the \nue CC inclusive
cross-section at the GeV scale~\cite{Blietschau:879197}, and there are currently no
\nue differential cross-section results as a function of the electron kinematics.
Theoretical differences are expected between \nue and \num
cross-sections~\cite{Day:2012gb}, and measuring these with data is critical to
understand the systematic uncertainties related to the search for CP violation in
the lepton sector. The uncertainty in \nue cross-sections will become increasingly
important in future oscillation experiments as statistical and other systematic
uncertainties are reduced.

In this Letter we present the first \nue CC inclusive differential
cross-section measurements for neutrinos with energy ${\sim}1$~GeV as a function of
the electron momentum (\elemom), electron scattering angle (\elecos)
and the four-momentum transfer of the interaction (\qq). The total flux-averaged
CC inclusive cross-section is also presented.

%% file: T2K-setup.tex
{\it T2K Experiment}\textemdash
T2K~\cite{Abe:2011ks} operates from the J-PARC facility in Tokai, Japan. A muon neutrino beam is
produced from the decay of charged pions and kaons generated by 30~GeV proton
collisions on a graphite target and focused by three magnetic horns. Downstream of the
horns is the decay volume, 96 meters in length, followed by the beam dump and muon
monitors~(MUMON~\cite{Matsuoka2010591}). The neutrino beam illuminates an on-axis
near detector (INGRID~\cite{Otani2010368}), an off-axis near detector
(ND280) and an off-axis far detector
(Super-Kamiokande~\cite{Fukuda2003418}). The off-axis detectors are positioned
at an angle of 2.5$^\circ$ relative to the beam axis direction. The near detectors are
located 280 meters from the target and are used to determine the neutrino beam
direction, spectrum, and composition before oscillations, and to measure neutrino
cross-sections. Super-Kamiokande, a 50~kt water Cherenkov
detector situated 295~km away, is used to detect the neutrinos after oscillation.

ND280 is a magnetized multi-purpose detector designed to
measure interactions of both \num and \nue from the T2K beam before oscillations.
It is composed of a number of subdetectors installed inside the refurbished
UA1/NOMAD magnet, which provides a magnetic field of 0.2~T. The central subdetectors
form a tracking detector, composed of two fine-grained scintillator detectors
(FGDs~\cite{Amaudruz:2012pe}) and three time projection chambers
(TPCs~\cite{Abgrall:2010hi}). The FGDs are used as the target for the neutrino
interactions, and while the upstream FGD (FGD1) is composed solely of scintillator
bars, the downstream FGD (FGD2) also contains water layers. Upstream of
the tracker
is a $\pi^0$ detector (P0D~\cite{Assylbekov201248}), explicitly built to measure
neutrino interactions with a $\pi^0$ in the final state. The tracker
and P0D are surrounded by a set of electromagnetic calorimeters
(ECals~\cite{Allan:2013ofa}), and the magnet yokes are instrumented with side muon
range detectors (SMRDs~\cite{Aoki:2012mf}) to track high angle muons.

The results presented here are based on data taken from January 2010 to May 2013.
During this period
the proton beam power has steadily increased and reached 220~kW continuous operation
with a world record of $1.2 \times 10^{14}$ protons per pulse. The physics-quality
data for this analysis corresponds to a total of $5.90 \times 10^{20}$ protons on
target (\pot).

%% file: neutrino_beam.tex
{\it Neutrino Beam Flux}\textemdash 
The neutrino beam flux~\cite{PhysRevD.87.012001} is predicted by modeling 
interactions of the primary beam protons with a graphite target using the FLUKA2008 
package~\cite{Ferrari:2005zk} and external hadron production data from the CERN 
NA61/SHINE experiment~\cite{Abgrall:2011ae, Abgrall:2011ts}.
GEANT3~\cite{GEANT3} with GCALOR~\cite{GCALOR} is used to simulate the propagation 
of secondary and tertiary pions and kaons, and their decays into neutrinos.
Decays of kaons and muons, in the decay volume, create the approximately 1\% \nue
component of the beam. Muon decays are the dominant source of \nue with energies 
below 1~GeV, with higher energy neutrinos produced by kaon decays.

The neutrino flux uncertainties are dominated by hadron production 
uncertainties, with contributions from the neutrino beam direction 
and the proton beam uncertainties. The neutrino beam direction---monitored indirectly 
by MUMON on a spill-by-spill basis, and directly by INGRID~\cite{Abe2012}---has 
been well within the required $\pm$1~mrad during the full run period. The 
neutrino interaction rate per \pot~has also been measured by 
INGRID, and is stable within 0.7\%. The total systematic uncertainty on the \nue flux is 13\% at the mean \nue energy (1.3~GeV).

%% file: data-selection.tex
{\it Selection of Electron Neutrino Interactions in ND280}\textemdash
Full details of the event selections can be found in Ref.~\cite{Abe:2014usb},
where the only difference is that in this analysis only interactions in FGD1 are
selected, rather than FGD1 and FGD2. This is so that interactions on water in FGD2
are not included.

\begin{figure}[t]
\includegraphics[width=0.99\linewidth]{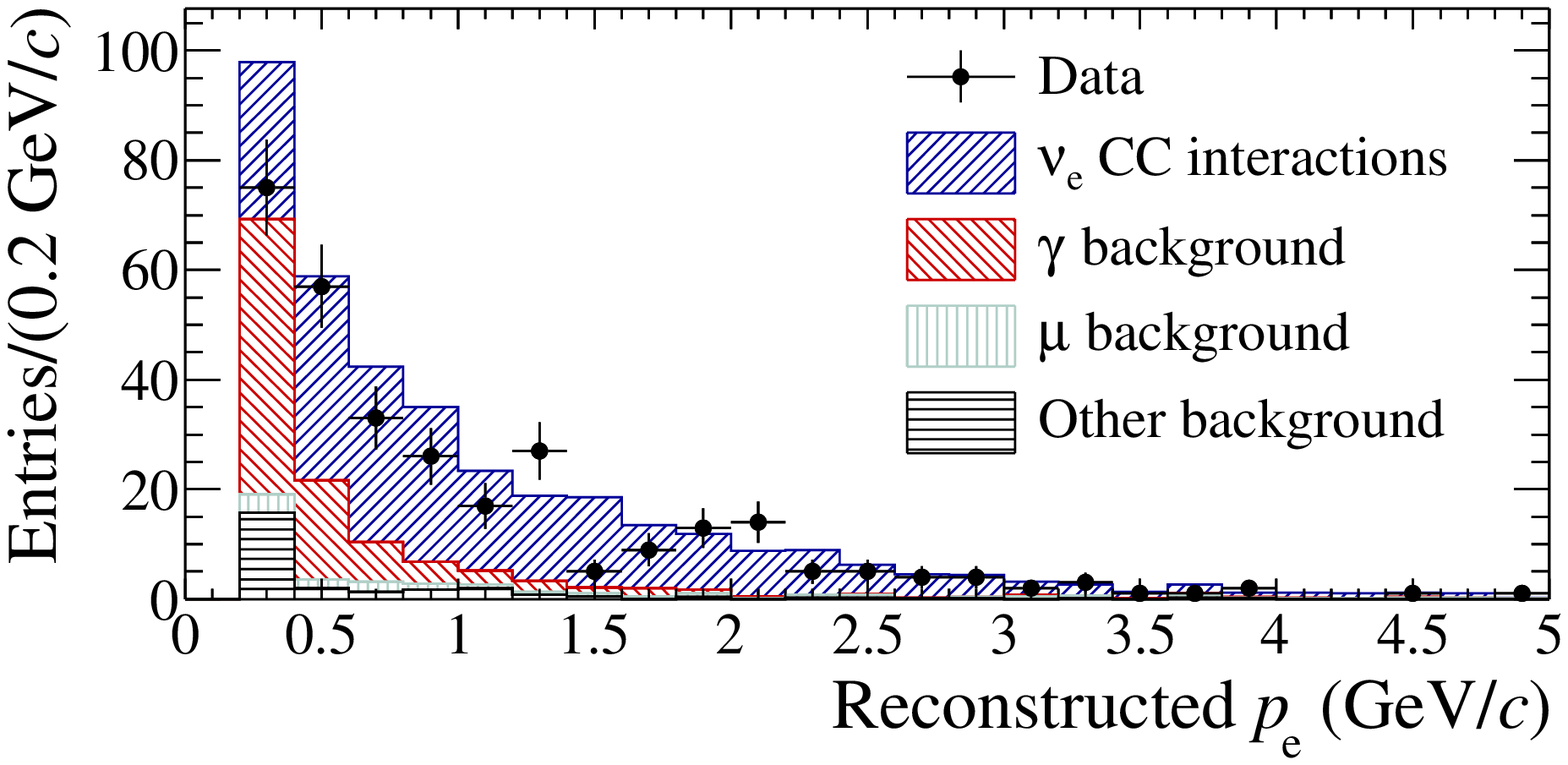}\\
\includegraphics[width=0.99\linewidth]{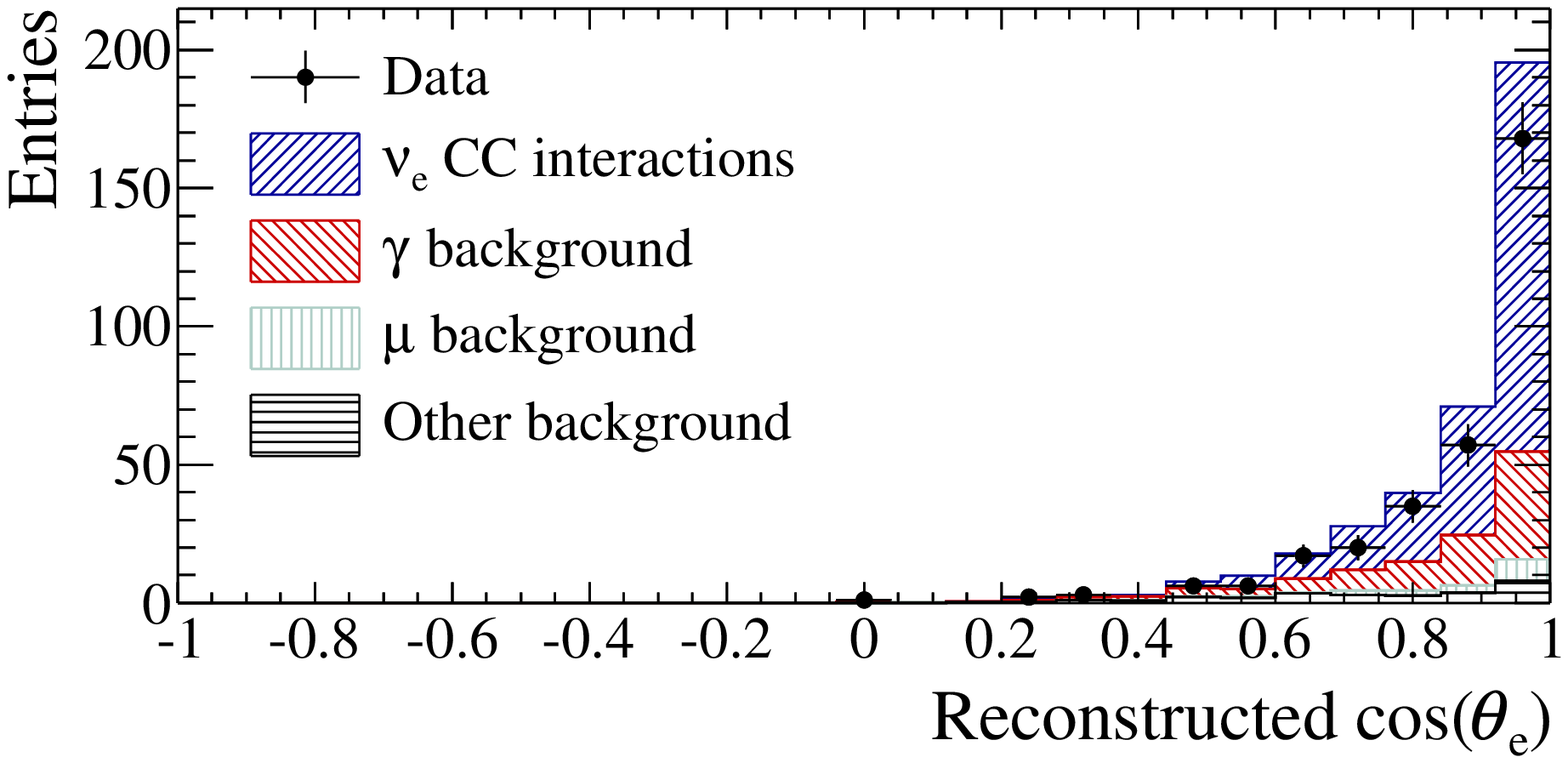}\\
\includegraphics[width=0.99\linewidth]{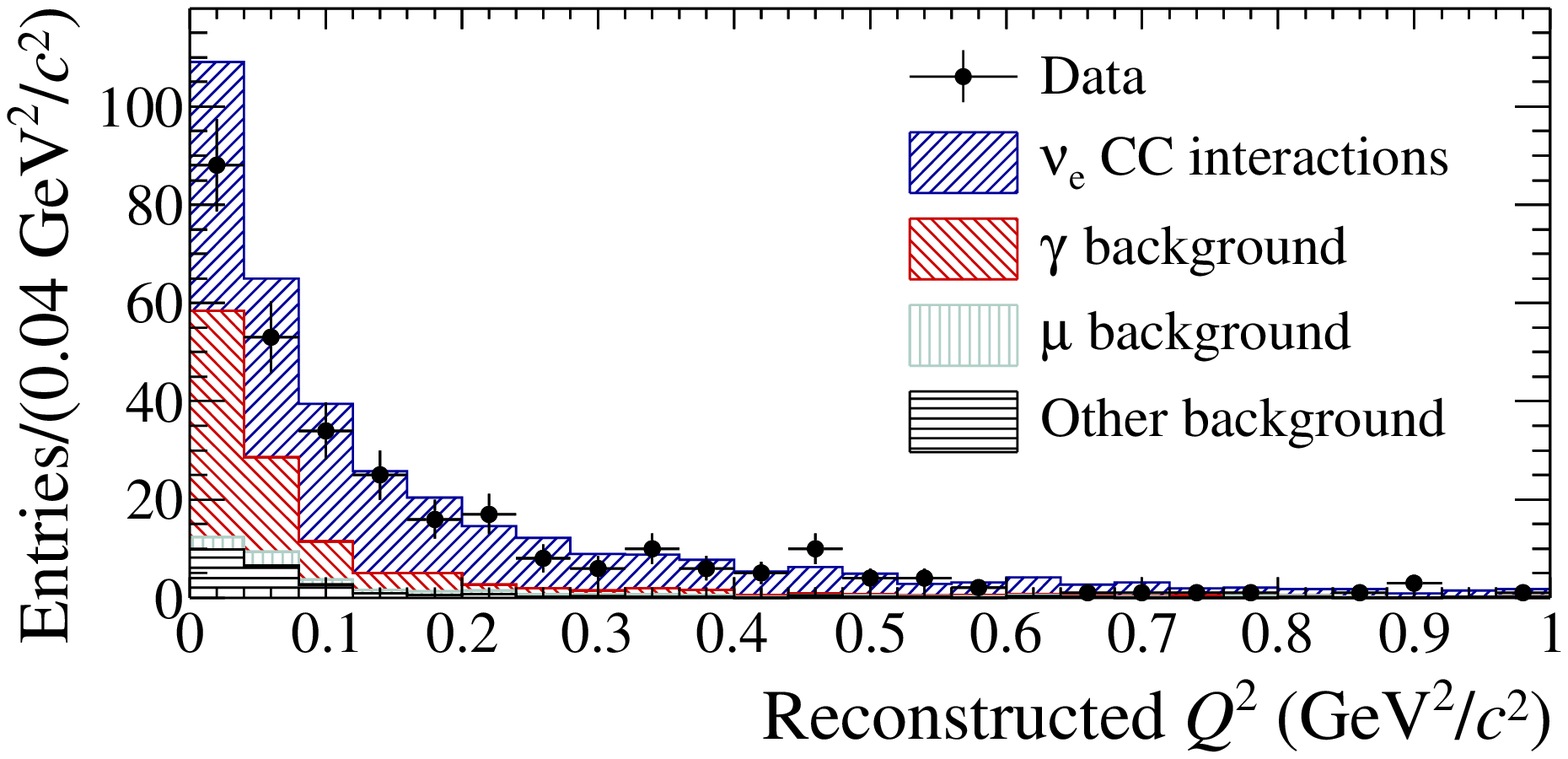}
\caption{\label{fig:nue_sel}Reconstructed \elemom (top), \elecos (middle) and \qq (bottom) distributions of \nue event candidates. The NEUT Monte Carlo prediction is separated into the \nue CC interaction signal, background from $\gamma\rightarrow\epem$ conversions, background from \mun tracks and all other backgrounds. The last bins in the top and bottom plots do not include the overflow of events.}
\end{figure}

Electron neutrino interactions are selected using the highest momentum
negative track starting inside the fiducial volume of FGD1. To reduce the large
background from \num charged current interactions, electron particle identification
criteria are applied using TPC $dE/dx$ and ECal shape and energy measurements.
These remove 99.9\% of \mun tracks, and although a clean
sample of \en is selected, 62.4\% of events are from photons which produce
\epem pairs in FGD1. This $\gamma$ background is reduced by searching for
a positron and applying an invariant mass cut, and vetoing on activity in TPC1, the P0D,
and ECals upstream of FGD1. After this procedure, 315 \nue CC interaction candidates
are selected, with an expected purity of 65\%. The reconstructed momentum,
scattering angle, and \qq distributions are shown in Fig.~\ref{fig:nue_sel}, and
compared to the prediction from the NEUT neutrino interaction generator~\cite{NEUT}.
\qq is reconstructed assuming
CC quasi-elastic (CCQE) kinematics~\cite{Abe:2013xua}, with a stationary target nucleon and 25~MeV
binding energy.

The background from $\gamma\rightarrow\epem$ conversions in the \nue sample is
23\%, 70\% of which are from neutrinos interactions outside the FGD1 fiducial
volume. A control sample, referred to as the $\gamma$ sample, is used to constrain
this, and is selected by finding electron-positron pairs that enter
the TPC and that have a low invariant mass. The data shows a deficit at low momentum
in both the \nue and $\gamma$ samples. This deficit is also visible in Ref.~\cite{Abe:2014usb}, which selects events in FGD2 as well as FGD1.

%% file: unfolding.tex
{\it Unfolding method}\textemdash
The Bayesian technique by d'Agostini~\cite{DAgostini1995487} is used to unfold
from the measured reconstructed distributions to the underlying true distributions.
For each observable, the true (reconstructed) bins are denoted by \tk (\rj). There
are \nt (\nr) true (reconstructed) bins in total.
Bayes' theorem is used to generate the unsmearing matrix
\begin{equation}
\label{Eqn:Bayes}
\unsmear = \frac{\smear\priorm}{\sum\limits_{\alpha=1}^{\nt}\smeara\priorma},
\end{equation}
where \smear is the smearing matrix and \priorm is the Monte Carlo (MC) prior probability of
finding a signal event in true bin \tk.
Given a dataset \meas, the estimated number of events in each true bin is given by
\begin{equation}
\label{Eqn:Unfold}
\initestmp = \frac{1}{\efftk}\sum_{j=1}^{\nr}\unsmear(\meas-\bkg),
\end{equation}
where \bkg is the number of background events that were selected and \efftk is the
efficiency of detecting a signal event in bin \tk.
The unfolding is performed separately for each variable. For defining the true bin of each interaction, the true momentum of the electron, true angle of the electron, and true \qq of the interaction from the generator are used for \elemom, \elecos and \qq, respectively. The NEUT neutrino generator is used for the unfolding results presented in this Letter.

The Bayesian unfolding technique was also used in Ref.~\cite{CCinc} for measuring
the \num CC inclusive cross-section with ND280. The main difference in the unfolding
method for this analysis is that
the MC background prediction, \bkg, is estimated using the $\gamma$
sample.
Specifically, the background from
neutrino interactions occurring outside of the fiducial volume (out-of-fiducial events) is
re-weighted based on the $\gamma$ sample data. This choice is made as the
systematic uncertainties relating to in-fiducial events have been well-studied,
30\% of the out-of-fiducial events
are on heavy targets (iron and lead) and 66\% are from interaction channels on which there are
large uncertainties in the modeling (deep inelastic scattering and neutral current
interactions).
The MC prediction of events in the fiducial volume is subtracted
from the $\gamma$ sample data, and the data/MC ratio of the out-of-fiducial events is then
computed in $(\elemom, \elecos)$ bins. The out-of-fiducial component of the \nue sample is re-weighted
based on this data/MC ratio distribution. The two-dimensional re-weighting
scheme is chosen as the \nue and $\gamma$ samples preferentially select photons
from different origins: the $\gamma$ sample requires both the \ep and \en to be
reconstructed, so preferentially selects higher-energy and more forwards-going
photons.

The effect of systematic uncertainties on the cross-section measurements are
computed using the same covariance matrix
method as in Ref.~\cite{CCinc}. Separate covariance matrices are computed for the data statistics, the MC
statistics, detector systematics, flux and cross-section systematics, and out-of-fiducial
systematics. One thousand toy experiments are performed to generate each matrix, and
each experiment simultaneously affects both the \nue and $\gamma$ samples.

The data statistical uncertainty is evaluated by varying the contents of each data bin according to Poisson statistics.
The MC statistical uncertainty is evaluated by separately varying the \nue, the
in-fiducial background and the out-of-fiducial background components
according to Poisson statistics.
Detector systematics are studied by varying parameters such as the
momentum resolution, and propagating the effect to the selection.
The TPC, FGD, ECal and external interaction uncertainties are described in detail
in Ref.~\cite{Abe:2014usb}.
The uncertainty on the FGD mass is 0.67\%~\cite{CCinc}.
The flux and cross-section uncertainties are also described in
Ref.~\cite{Abe:2014usb}. The flux uncertainties are based on beamline measurements
and hadron production data.
The cross-section uncertainties, including neutrino-nucleon, nuclear modeling, pion production and final state interaction uncertainties are constrained using external data and comparisons between different nuclear models~\cite{Abe:2013xua}; these uncertainties affect signal efficiencies and background spectra.

Due to the discrepancy between data and MC in the $\gamma$ sample,
conservatively an extra systematic is applied to the out-of-fiducial re-weighting
in addition to the
statistical uncertainty of the $\gamma$ sample. If the
re-weighting factor in a given bin is $\alpha$, then the correction
is modeled as a Gaussian with mean $\alpha$ and width $\alpha/3$.

%% file: results.tex
{\it Cross-section results}\textemdash
The signal for this analysis is all \nue CC interactions occurring in the FGD1
fiducial volume. FGD1 is composed of carbon (86.1\% by mass), hydrogen (7.4\%),
oxygen (3.7\%), titanium (1.7\%), silicon (1.0\%) and nitrogen (0.1\%).
The analysis measures the flux-averaged differential \nue CC inclusive
cross-section, and for bin \tk of variable $X$, this is given by
\begin{equation}
\label{Eqn:XSec}
\left\langle\frac{\partial\langle\sigma\rangle_\phi}{\partial X}\right\rangle_{\tk} = \frac{N_{\tk}}{\Delta X_{\tk}T\phi},
\end{equation}
where $X$ is either \elemom, \elecos or \qq, $\Delta X_{\tk}$ is the width of the bin, $N_{\tk}$ is the total number of
signal events in the bin, $T$ is the number
of target nucleons ($5.5\times10^{29}$~\cite{CCinc}), $\phi$ is the total integrated
flux ($1.35\times10^{11}~\mathrm{cm}^{-2}$), and $\langle\cdots\rangle_\phi$ indicates
that the quantity is averaged over the flux.

The total flux averaged cross-section per nucleon is computed by summing
over all $X$ bins, as
\begin{equation}
\label{Eqn:XSecTotal}
\fluxav = \frac{\sum_{k=1}^{\nt}N_{\tk}}{T\phi}.
\end{equation}

For comparison, differential and total flux-averaged cross-section predictions are
computed using the NEUT (version 5.1.4.2) and GENIE
(version 2.6.4~\cite{Andreopoulos:2009rq}) generators.

\begin{figure}[t]
\includegraphics[width=0.99\linewidth]{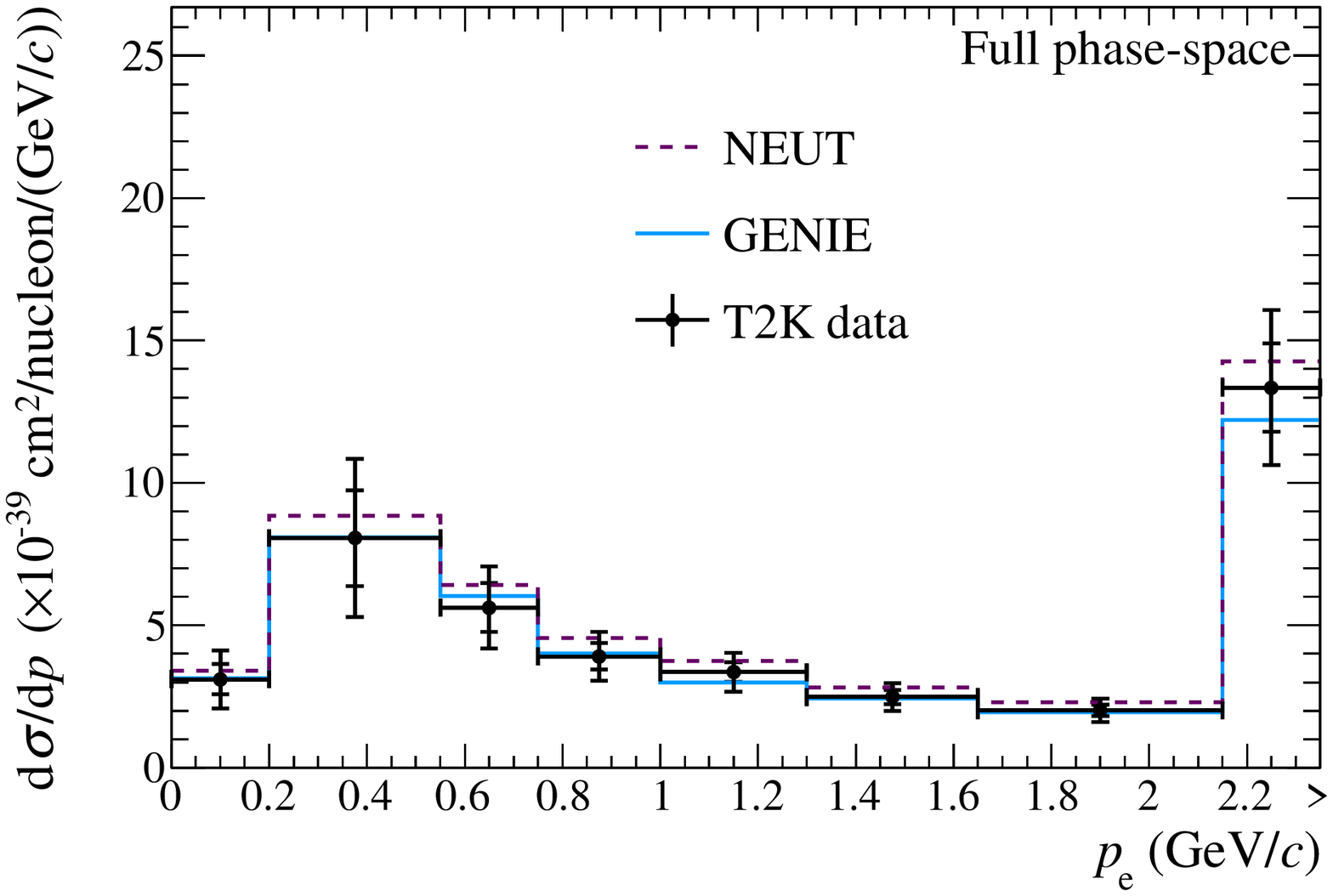}\\
\includegraphics[width=0.99\linewidth]{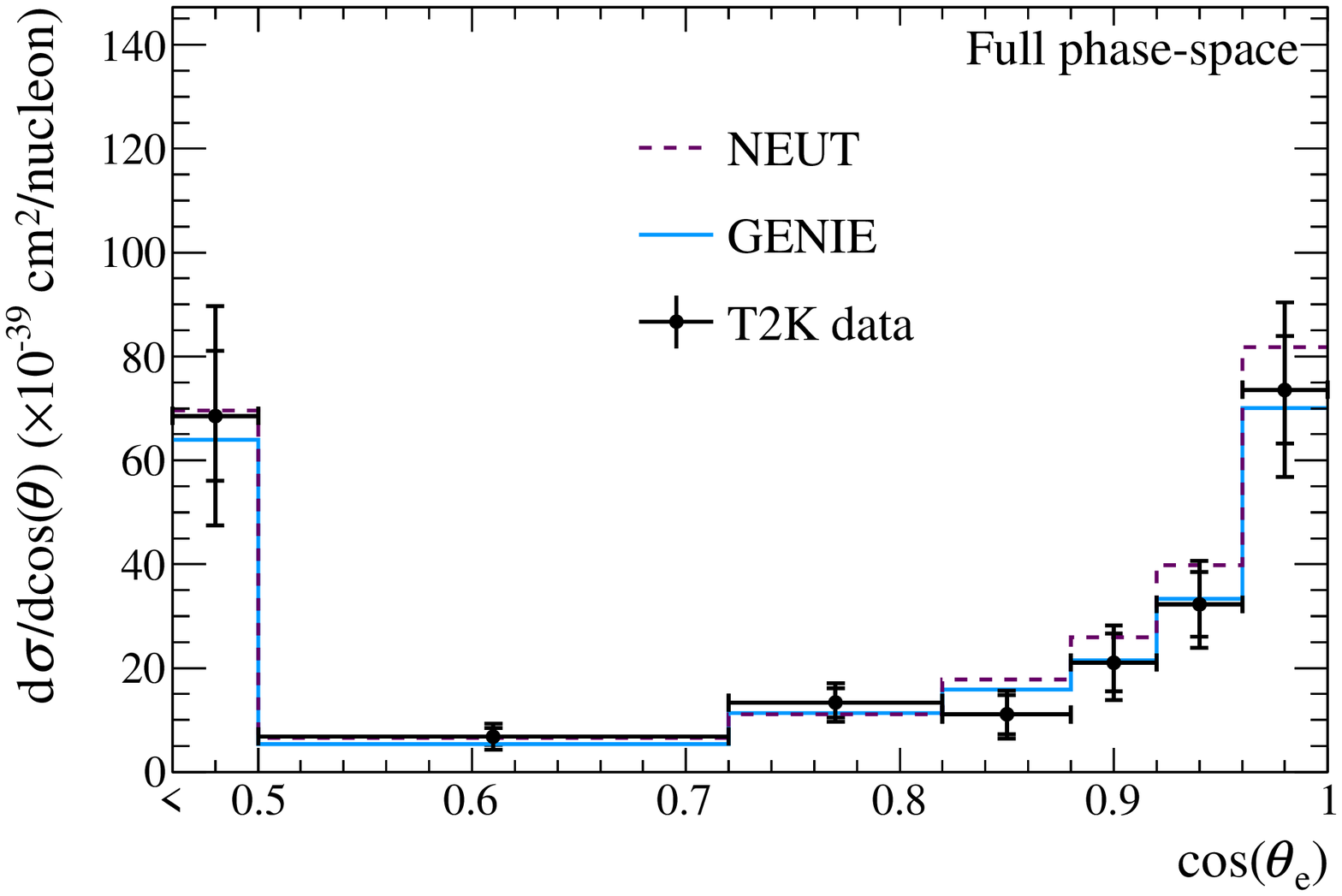}\\
\includegraphics[width=0.99\linewidth]{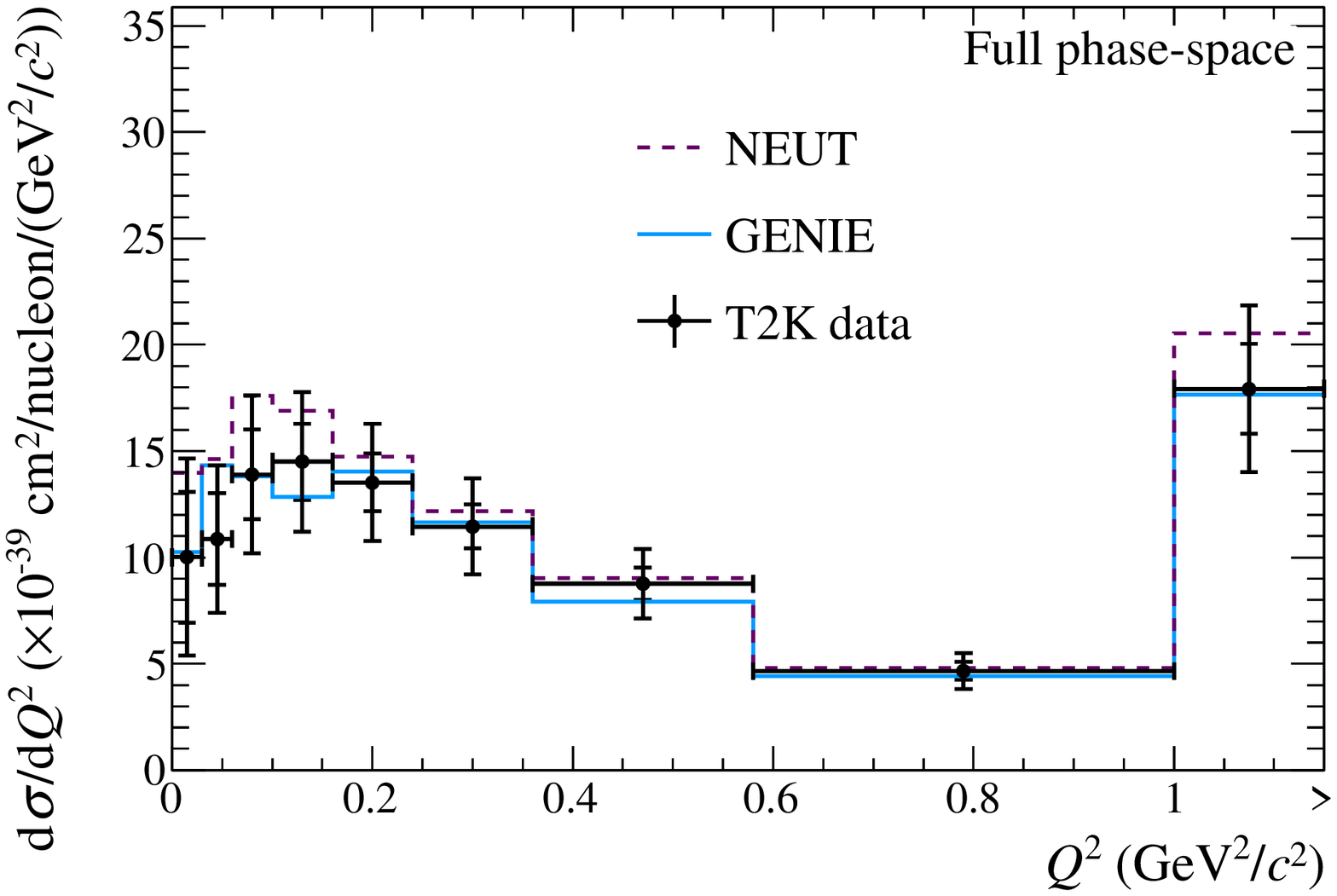}
\caption{\label{fig:results_diff}Unfolded \nue CC inclusive differential
cross-sections as a function of \elemom (top), \elecos (middle) and \qq (bottom).
The inner (outer) error bars show the statistical (total) uncertainty on the data.
The dashed (solid) line shows the NEUT (GENIE) prediction. Overflow (underflow)
bins are indicated by $>$ ($<$) labels, and are normalized to the width shown.}
\end{figure}

Fig.~\ref{fig:results_diff} shows the unfolded differential cross-section results
as a function of \elemom, \elecos and \qq. The data agrees with both NEUT and
GENIE, although a deficit is seen at low \qq compared to NEUT. The biggest
differences between NEUT and GENIE at low \qq are caused by the different values of
$M_A^{QE}$ chosen for CCQE interactions, and different CC coherent interaction models.

\begin{figure}
\includegraphics[width=0.99\linewidth]{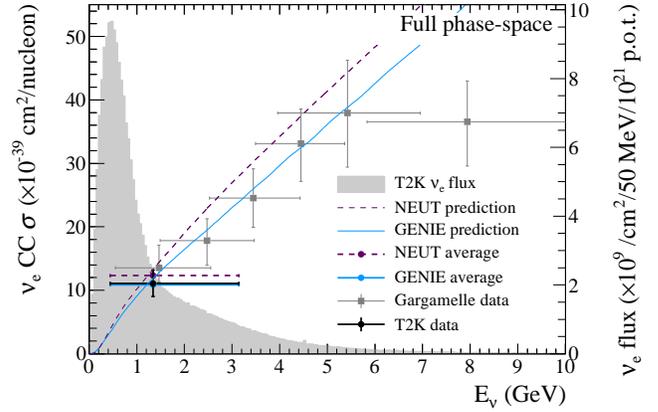}
\caption{\label{fig:results_av}Total \nue CC inclusive cross-section when unfolding
through \qq. The T2K data point is placed at the \nue flux mean energy. The vertical
error represents the total uncertainty, and the horizontal bar represents 68\% of the flux each side of the mean. The T2K flux
distribution is shown in grey.
The NEUT and GENIE predictions are the total \nue CC inclusive
predictions as a function of neutrino energy. The NEUT and GENIE averages are the
flux-averaged predictions. The Gargamelle data is taken from
Ref.~\cite{Blietschau:879197}.}
\end{figure}

The total flux-averaged cross-section when unfolding through \qq is
$\fluxav=1.11\pm0.09~(\mathrm{stat})\pm0.18~(\mathrm{syst})\times10^{-38}~\mathrm{cm}^{2}/\mathrm{nucleon}$, which agrees
with both the NEUT prediction of
$1.23\times10^{-38}~\mathrm{cm}^{2}/\mathrm{nucleon}$ and the GENIE prediction of
$1.08\times10^{-38}~\mathrm{cm}^{2}/\mathrm{nucleon}$. The result is shown in
Fig.~\ref{fig:results_av}, along with the Gargamelle data from 1978~\cite{Blietschau:879197}. The results when unfolding through the other
variables agree at the percent level. The dominant systematic uncertainties on this
result are the flux (12.9\%) and detector systematics (8.4\%), with all other
systematics giving a 6.1\% uncertainty when added in quadrature. The uncertainty
from re-weighting the out-of-fiducial background is 2.1\%.

\begin{figure}
\includegraphics[width=0.99\linewidth]{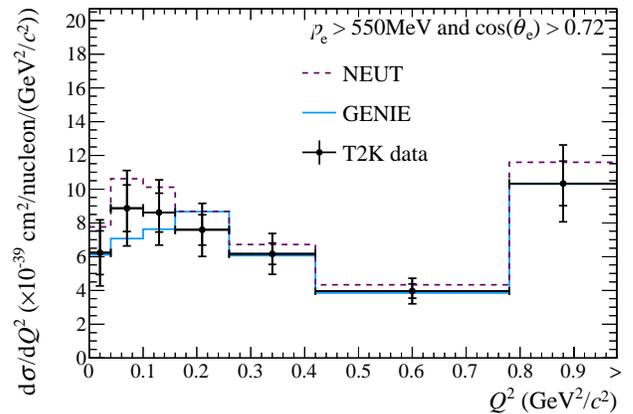}
\caption{\label{fig:results_diff_restrict}Unfolded \nue CC inclusive differential
cross-section as a function of \qq, when only electrons with
$\elemom>550~\mathrm{MeV}$ and $\elecos>0.72$ are considered.
The inner (outer) error bars show the statistical (total) uncertainty on the data.
The dashed (solid) line shows the NEUT (GENIE) prediction. Overflow (underflow)
bins are indicated by $>$ ($<$) labels, and are normalized to the width shown.}
\end{figure}

An important aspect of the Bayesian unfolding approach is that it allows a
reconstructed distribution to be unfolded into regions that it is not sensitive to.
This analysis has poor reconstruction efficiency for low momentum, backwards going,
or high angle electrons. This adds model dependency since the NEUT generator must
predict these poorly determined regions.
For this reason, a second result is presented, in which
only events with $\elemom>550~\mathrm{MeV}$ and $\elecos>0.72$ are considered. In
this ``reduced phase-space" result, no attempt is made to unfold into regions of
low detector efficiency.
The unfolded \qq differential cross-section result for this reduced phase-space is
shown in Fig.~\ref{fig:results_diff_restrict}.

%% file: conclusion.tex
{\it Conclusion}\textemdash
Understanding differences between \nue and \num cross-sections is vital as long
baseline oscillation experiments search for CP violation in the lepton sector.
The T2K off-axis near detector, ND280, has been used to extract \nue CC
inclusive flux-averaged differential cross-sections as a function of \elemom,
\elecos and \qq, and they are found to agree with both the NEUT and GENIE
neutrino interaction generator predictions. These are the first ever \nue
differential cross-section measurements at the GeV-scale. The total \nue CC
inclusive flux-averaged cross-section is found to be
$1.11\pm0.20\times10^{-38}~\mathrm{cm}^{2}/\mathrm{nucleon}$, which is also in
agreement with the NEUT and GENIE predictions. The data related to the measurement can be found in \cite{Abe:datarelease}. 

%% file: acknowledge.tex
We thank the J-PARC staff for superb accelerator performance and the
CERN NA61 collaboration for providing valuable particle production
data. We acknowledge the support of MEXT, Japan; NSERC, NRC, and CFI,
Canada; CEA and CNRS/IN2P3, France; DFG, Germany; INFN, Italy;
National Science Centre (NCN), Poland; RAS, RFBR, and MES,
Russia; MICINN and CPAN, Spain; SNSF and SER, Switzerland; STFC, UK;
and DOE, USA. We also thank CERN for the UA1/NOMAD magnet, DESY for
the HERA-B magnet mover system, NII for SINET4, the WestGrid and
SciNet consortia in Compute Canada, GridPP, UK, and the University of
Oxford Advanced Research Computing (ARC) facility.
In addition
participation of individual researchers and institutions has been
further supported by funds from: ERC (FP7), EU; JSPS, Japan; Royal
Society, UK; DOE Early Career program, USA.